\documentclass{article}
\usepackage{amsmath,amstext,amsfonts}
\usepackage[dvips]{graphicx}
\usepackage{subfigure}
\providecommand{\keywords}[1]{\textbf{\textit{Keywords-}} #1}
\providecommand{\subjclass}[1]{\textbf{\textit{MSC 2000-}} #1}
\newcommand{\goodgap}{%
\hspace{\subfigtopskip}%
\hspace{\subfigbottomskip}
}
\begin{document}
\title{Circular time-like geodesics around a charged spherically symmetric dilaton black hole}
\author{Cristina Blaga\\Faculty of Mathematics and Computer Sciences\\ Babe\c{s}-Bolyai University of Cluj-Napoca\\ 1 Kog\u{a}lniceanu Street, 400084 Cluj-Napoca, Romania} 
\maketitle
%\def\AMSCLASSIFICATION{83C10, 83C20, 83C57}% 
%
%\def\KEYWORDS{circular time-like geodesics, dilaton black holes, effective potential}% 
%\keywords{circular time-like geodesics, dilaton black holes, effective potential}
%\subjclass[2000]{83C10, 83C20, 83C57}
\begin{abstract}
In this note we examine the circular time-like geodesics near a spherically symmetric dilaton black hole, described using the exact solution for a static charged black hole found by Gibbons and Maeda and, independently, by Garfinkle, Horowitz and Strominger. The existence and stability of the circular orbits are analysed using the effective potential of a free material test particle moving on time-like geodesic near this black hole. We determine the radius of the innermost stable circular orbit, the radius of the shortest circular orbit and compare our results with those obtained by other authors for specific values of the parameters involved in our analysis.
\end{abstract}



\section{Introduction}
In classical general relativity, the geometry of the space-time near a charged black hole is described by the Reissner-Nordstr\o m solution. In low-energy string theory the solution for the static charged black hole was found by Gibbons and Maeda \cite{CBlaga-gm} and, independently, three years later, by Garfinkle, Horowitz and Strominger \cite{CBlaga-ghs}. In literature this solution is known as Gibbons-Maeda-Garfinkle-Horowitz-Strominger (GMGHS) black hole. In 1993, using a solution generating technique, Horowitz \cite{CBlaga-hor} derived a metric that fulfills Einstein-Maxwell dilaton field equations. He showed that this new solution is the static spherically symmetric solution for a charged massless dilaton obtained earlier by Gibbons and Maeda, respectively Garfinkle, Horowitz and Strominger. The line element for this metric is:  
\begin{equation}\label{CBlaga-metr}
ds^2=-\left(1-\frac{2M}{r}\right) dt^2 + \left(1-\frac{2M}{r} \right)^{-1} d r^2 + r \left(r-\frac{Q^2}{M}\right) (d \theta^2 + \sin^2 \theta \, d \varphi^2),
\end{equation} 
where $Q$ is related to the electrical charge of the black hole and $M$ to its mass. For $Q^2 < 2 M^2$ this black hole has an event horizon. In the case $Q^2 = 2 M^2$ the area of the horizon shrinks to zero and the solution describes a naked singularity. This case is named extremal GMGHS black hole.         

The paper is structured as follows: in section 2 we derive the geodesics equations. In the third section we analyse the effective potential for time-like geodesics. The potential depends on several variables: radial coordinate $r$, angular momentum of the particle $L$ and two parameters $M$ and $Q$ related to mass, respectively electrical charge of black hole. When we mention derivatives, we mean derivatives in respect to $r/M$, variable proportional to the radial coordinate. The circular time-like geodesics correspond to extrema of the effective potential, therefore we examine the existence of the zeros of its first derivative. Further we discuss the stability of the circular orbits and determine the radius of the innermost stable orbit and the radius of the shortest circular orbit. In the end, we compare our results with those acquired by other authors for Schwarzschild black hole and extremal GMGHS black hole, cases obtained for specific values of the parameters involved in this analysis. The study of time-like geodesics in a Schwarzschild space-time was made by Chandrasekhar, \cite{CBlaga-cha}. For an extremal GMGHS black hole the innermost stable circular orbit was found by Pradhan in \cite{CBlaga-ppp12}.        
\section{The geodesics equations}
A free test particle moves around a black hole along a time-like geodesics. We can derive the geodesics equations directly, using the Hamilton Jacobi theory (see Blaga and Blaga \cite{CBlaga-bb}) or writing the Euler-Lagrange equations (see Fernando \cite{CBlaga-f12}). We will follow the last approach described in detail by Chandrasekhar (see \cite{CBlaga-cha}). The Lagrangian for the metric (\ref{CBlaga-metr}) is: 
\begin{equation}\label{CBlaga-lagr}
2 \mathcal{L} = -\left(1-\frac{2M}{r}\right) \dot{t}^2 + \left(1-\frac{2M}{r} \right)^{-1} \dot{r}^2 + r \left(r-\frac{Q^2}{M}\right) \left(\dot{\theta}^2  + \sin^2 \theta \,\dot{\varphi}^2 \right) 
\end{equation}
where the dot means the differentiation with respect to $\tau$ - an affine parameter along the geodesic. We will choose this parameter so that $2 \mathcal{L}=-1$ on a time-like geodesics, $2 \mathcal{L}=0$ on a null geodesics and $2 \mathcal{L}=1$ on a space-like geodesics.  

The coordinates $t$ and $\varphi$ are cyclic, therefore the motion admits two integrals. The first integral of motion is 
\begin{equation}\label{CBlaga-ien}
\left(1-\frac{2M}{r}\right) \dot{t} = \text{constant} = E (\text{say})
\end{equation}
the energy integral. The constant $E$ is the total energy of the particle. 

The second integral of motion is
\begin{equation}\label{CBlaga-imc}
2 \, \sin^2\theta \, \cdot r \left(1-\frac{2M}{r}\right) \dot{\varphi} = \text{constant}
\end{equation}
the integral of angular momentum. 

From the Euler-Lagrange equation for the coordinate $\theta$ we get
\begin{equation}\label{CBlaga-ecteta}
\frac{d}{d \tau}\left[ r \left( r - \frac{Q^2}{M}  \right) \dot{\theta} \right] = r \left( r - \frac{Q^2}{M}  \right) \sin \theta \cos \theta \cdot \dot{\varphi}^2\,.
\end{equation}
If we consider that $\theta = \pi/2$ when $\dot{\theta}=0$, then from (\ref{CBlaga-ecteta}) it follows that $\ddot{\theta} = 0$ and $\theta=\pi/2$ along the geodesic. Hence the geodesics are planar, like in Schwarzschild space-time or in Newtonian gravitational field. If $\theta=\pi/2$, then the angular momentum integral (\ref{CBlaga-imc}) becomes: 
\begin{equation}
r \left(1-\frac{2M}{r}\right) \dot{\varphi} = L
\end{equation} 
where $L$ is the angular momentum about an axis normal at the plane in which the motion took place. 

The Euler-Lagrange equation corresponding to $r$ coordinate is complicated. Usually it is replaced with the condition of constancy of the Lagrangian, in which the integrals of motion are substituted. After some algebra, we obtain 
\begin{equation}\label{CBlaga-et}
\left( \frac{dr}{d \tau} \right)^2 + \left( 1 - \frac{2 M}{r} \right) \left(\frac{L^2}{r \left( r - \frac{Q^2}{M} \right)}  - \epsilon \right) = E^2
\end{equation}   
where $\epsilon=-1$ for time-like geodesics, $\epsilon=0$ for null geodesics and $\epsilon=+1$ for space-like geodesics. For the time-like geodesics, the parameter $\tau$ is the proper time of the particle describing the geodesics. If we compare the relation (\ref{CBlaga-et}) with the energy conservation law from the motion of a particle in the Newtonian gravitational field, we notice that the second term from the left-hand side of the relation (\ref{CBlaga-et}) could be interpreted as the effective potential. We introduce    
\begin{equation}\label{CBlaga-Veff}
V_{\text{eff}} = \left( 1 - \frac{2 M}{r} \right) \left(\frac{L^2}{r \left( r - \frac{Q^2}{M} \right)}  - \epsilon \right) 
\end{equation}
known as effective potential. In literature, sometimes the function (\ref{CBlaga-Veff}) is denoted by $V_{\text{eff}}^2$, because if the energy is positive, the motion is possible only if $E \geq \sqrt{V_{\text{eff}}}$. The points $r$ for which the total energy equals effective potential are known as radial turning points. 
\section{The time-like geodesics}
For time-like geodesics $\epsilon=-1$. With the notations $u=r/M$, $a=L^2/M^2$, $b=Q^2/M^2$, the effective potential (\ref{CBlaga-Veff}) becomes
\begin{equation}
V_{\text{eff}} = \left( 1 - \frac{2}{u} \right) \left(1 + \frac{a}{u \left( u - b \right)} \right) \,.
\end{equation}
The quantity $\mathit{q}=Q/M$ is the specific electrical charge. We notice that $a=\mathit{l}^2$ and $b=\mathit{q}^2$. Thus, in our study, we consider $a>0$ and $b \in \left[0,2 \right]$. The case $b=0$ corresponds to a Schwarzschild black hole and $b=2$ to an extremal GMGHS black hole. Outside the event horizon $u>2$. The effective potential $V_{\text{eff}}$ is not defined in $u=0$ or $u=b$, it is positive for $u \in (0,b) \cup [2, +\infty)$, negative for $u \in (b,2)$. It tends to $+ \infty$ for $u=0$ and $u=b$, it is zero at the event horizon and tends to $1$ when $u \rightarrow + \infty$.   

In the figure 1 we have plotted $\sqrt{V_{\text{eff}}}$ against $u=r/M$ for different values of $a$ and $b$. The motion is possible only if $E \geq \sqrt{V_{\text{eff}}}$. If $E = \sqrt{V_{\text{eff}}}$, then from (\ref{CBlaga-et}) it follows that $d r/ d \tau=0$ and the test particle might change its direction of motion. These points are called radial turning points. If $d r/d \tau=0$ and $d V_{\text{eff}}/d r=0$, then the particle describes a circular orbit. We notice that only for specific values of $a$ and $b$ the motion is bounded. If we compare the plots for different values of $b$ - parameter related to the electrical charge of the black hole - we notice that if $b$ increases, the energy of the particles reaching the event horizon increases (figure 1d).    
\begin{figure}
\centering
\subfigure{\includegraphics[width=.38\textwidth]{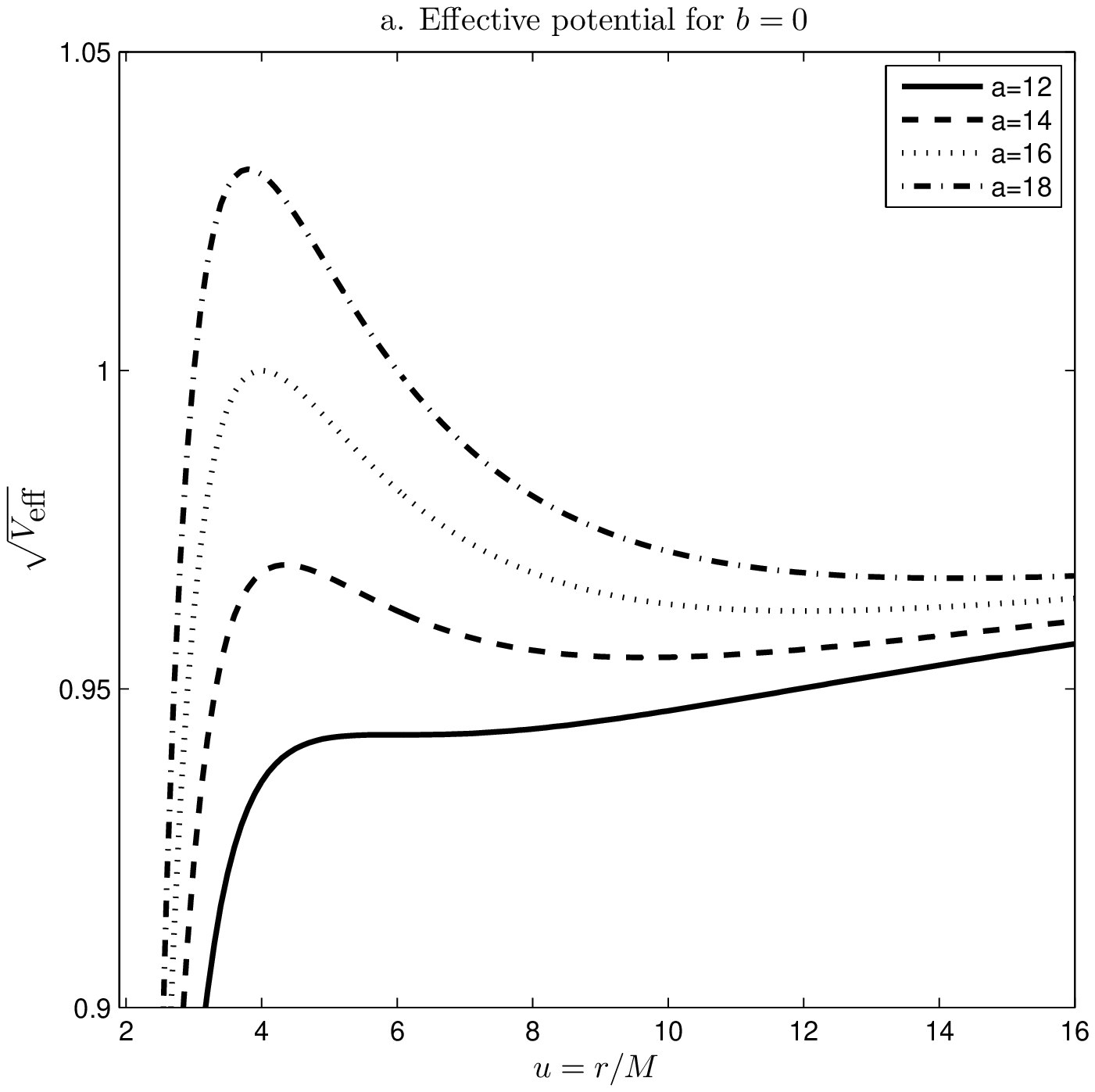}}\goodgap
\subfigure{\includegraphics[width=.38\textwidth]{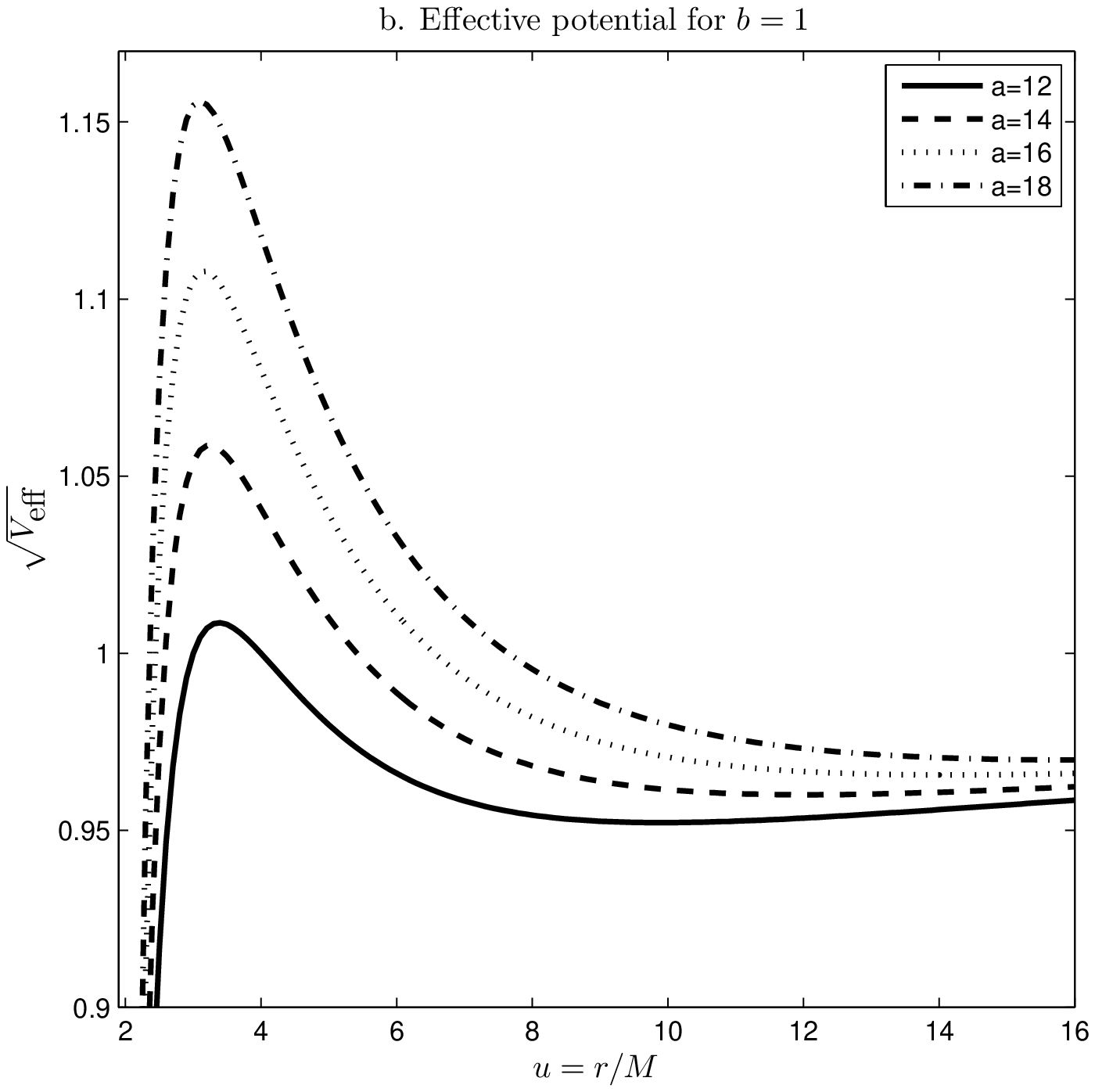}}\\
\subfigure{\includegraphics[width=.38\textwidth]{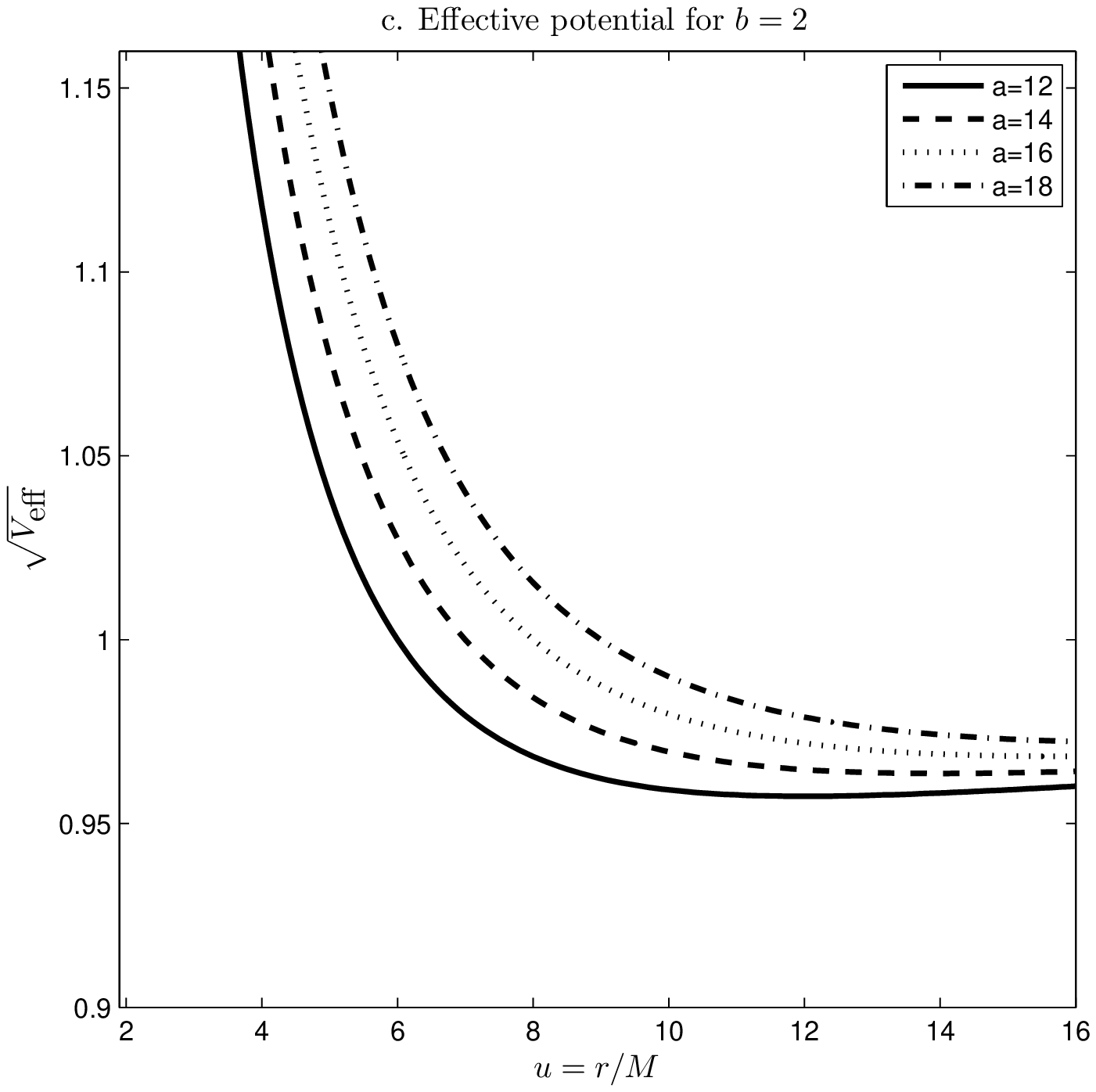}}\goodgap
\subfigure{\includegraphics[width=.38\textwidth]{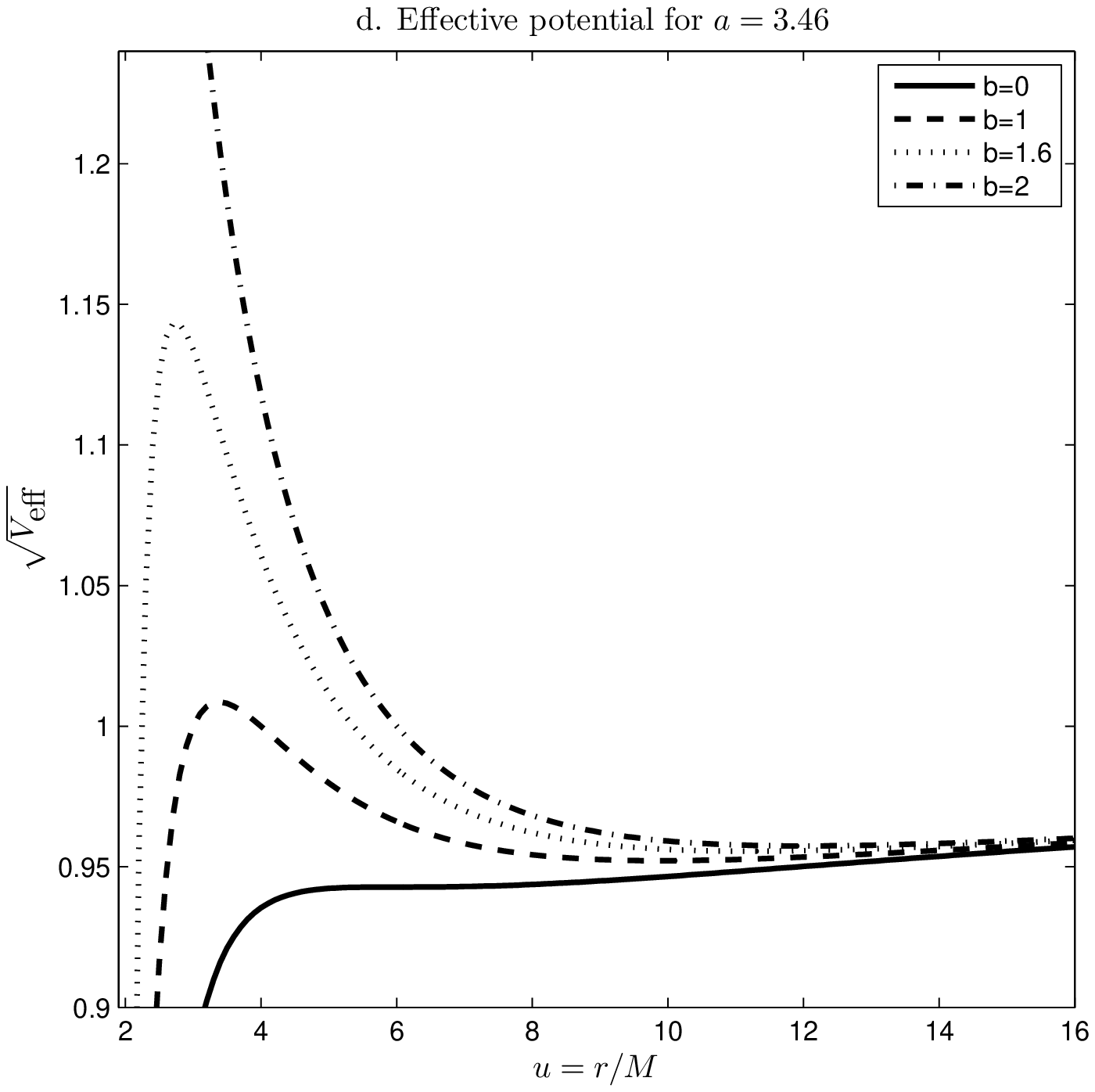}}
\caption{Effective potential for a Schwarzschild black hole ($b$=0), a dilaton black hole with $b=1$, an extremal GMGHS black hole ($b=2$) for different specific angular momentum $a = L^2/M^2$ and a dilaton black hole with different specific electrical charge $b = Q^2/M^2$ and $a=12$.}
\end{figure}
\subsection{Circular geodesics}
For circular orbits $d r/ d \tau = 0$ and $d V_{\text{eff}}/d r=0$. The new variable $u$ is proportional with $r$, therefore the last condition is equivalent with $d V_{\text{eff}}/ du=0$, which follow us to
\begin{equation}\label{CBlaga-cond}
{\frac {2\,{u}^{3}- 2\, \left(a +2\, b\right) {u}^{2}+ \left( 2\,{b}^{2}+6\,a+ab \right) u-4\,ab}{{u}^{3} \left( b - u \right) ^{2}}} = 0 \,.
\end{equation}
The circular orbits correspond to the roots of the numerator - a third degree polynomial in $u$. The nature of its roots depends on the sign of the discriminant
\begin{equation}\label{CBlaga-Delta}
\Delta = - \frac{a \, \left(b-2 \right) \, h(a,b)}{432} \,,
\end{equation} 
where 
\begin{equation}
h(a,b)=\left( b-18 \right) {a}^{3}+ \left( 6\,{b}^{2}-72\,b+216 \right) {a}^
{2}+ \left( 12\,{b}^{3}-72\,{b}^{2} \right) a+8\,{b}^{4}\,,
\end{equation} 
is a third degree polynomial in $a$. If $\Delta>0$, there will be one real root and two conjugate complex roots, if $\Delta=0$ there will be three real roots, at least two  of which being equal, while if $\Delta<0$, there will be three real and unequal roots. 

In our analysis $a>0$ and $0<b \leq 2$, hence the sign of the discriminant is the sign of the function $h$. If $b=2$ then $\Delta=0$ for all $a$. To determine the sign of function $h$ we will consider it as a function of variable $a$ and parameter $b$. The function $h(a,b)$ tends to $ + \infty$ if $a \rightarrow - \infty$ and $h(a,b) \rightarrow - \infty$ if $a \rightarrow + \infty$. The zeros of $h$ are the roots of the cubic equation in $a$. Its discriminant is
\begin{equation}
\bar{\Delta} = 186624\,{\frac {{b}^{4} \left( b - 2 \right) ^{2}}{ \left( b-18 \right) ^{4}}} \, \geq \, 0 \quad \forall \, b \in \mathbb{R} \setminus \{18\} \,.
\end{equation}    
Hence the function $h$ has a real root, except the cases $b=0$ and $b=2$. If $b=0$ the roots are $a=0$ - double root and $a=12$ a single root and if $b=2$ then $a=2$ is a triple root. The real root of the discriminant $\Delta$ is 
\begin{equation}
\bar{a}= 2\, {\frac {3 \, \left( b-6 \right) \sqrt[3]{8-4\,b}- 18 \, \sqrt [3]{2} \left( 2-b \right) ^{2/3} -\left( b-6
 \right) ^{2}}{b-18}} 
\end{equation}
We have plotted in figure 2 the function $\bar{a}=\bar{a}(b)$. We notice that $\bar{a}(b)$ decreases from 12 to 2, when $b$ increases from $0$ to $2$. 
\begin{figure}
\centering
\includegraphics[width=.5\textwidth]{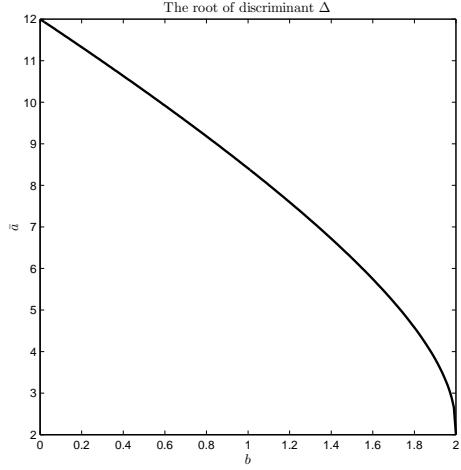}
\caption{The plot of function $\bar{a}=\bar{a}(b)$ - the real root of the discriminant $\Delta$ of the first derivative of the effective potential $V_{\text{eff}}$.}
\end{figure}
We conclude that if $a \in [ 0, \bar{a})$ then $d V_{\text{eff}}/d r=0$ has one real root. If $a=\bar{a}$ the derivative of the effective potential has two real roots (one single root and a double root) and if $a > \bar{a}$ then $d V_{\text{eff}}/d r=0$ has three real roots. 

Studying the relations between the coefficients and the roots of the third degree polynomial from the numerator of $d V_{\text{eff}}/d r$, we noticed that, in our case, the sum and the product of the roots are positive, thus the third order equation has at least one positive root. It corresponds to a circular orbit inside the event  horizon\footnote{This statement is easily proved if we study the variation of the effective potential or the relations between the coefficients and the roots of the third degree polynomial corresponding to the numerator of relation (\ref{CBlaga-cond}) in the variable $y=u-2$.}. The other roots of the numerator correspond to circular orbits outside the event horizon.        
\subsection{Stability of the circular geodesics}

For a given $b$, if $a>\bar{a}$ there will be two circular orbits outside the event  horizon. A circular orbit is stable if it locates a minimum of the effective potential and unstable if it locates a maximum of the effective potential curve. From the figure 1, we have noticed that outside the events horizon, the smaller root of the first derivative of the effective potential is unstable and the greater is the stable one. The radius of the stable orbit tends to infinity when $a$ increases endlessly for all $b$. The smallest value of the radius of the stable circular geodesics corresponds to the innermost stable circular orbit (ISCO). We have plotted it in the figure 3 with solid line. The radius of the ISCO decreases from $6M$ - at $b=0$ Schwarzschild black hole - to $2M$ at $b=2$ extremal GMGHS black hole. We have obtained the same values like other authors who derived the radius of the innermost stable orbit for these two extreme black holes: Schwarzschild black hole - Chandrasekhar in \cite{CBlaga-cha} and for an extremal GMGHS black hole - Pradhan in \cite{CBlaga-ppp12}. 

The limit of the radius of the unstable circular orbit when $a$ tends to infinity depends on $b$. This limit corresponds to the circular orbit with the smallest radius, the so called shortest circular orbit (SCO). It is plotted with dashed line in the figure 3 and decreases from $3$ to $2$ when $b$ increases from $0$ to $2$. The value of the shortest circular orbit for Schwarzschild black hole was obtained by Chandrasekhar in \cite{CBlaga-cha}. 
\begin{figure}
\centering
\includegraphics[width=.5\textwidth]{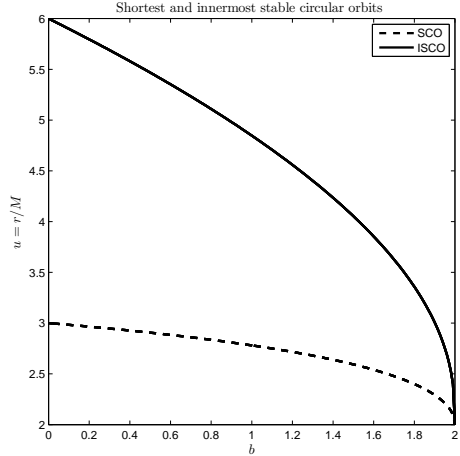}
\caption{The radii of shortest circular orbit (SCO) and the innermost stable circular orbit (ISCO) for charged spherical symmetric dilaton black holes. Plot of $u=r/M$ versus $b$.}
\end{figure}

In conclusion, for a given $b$, if $a \geq \bar{a}(b)$ there are circular time-like geodesics near the given charged spherical symmetric black hole. The radius of the unstable circular orbit is located between the radius of SCO and the radius of ISCO (corresponding to the given $b$), between the dashed and solid line in figure 3. The radius of the stable circular time-like geodesic is $\geq r_{ISCO}$. The radii of the circular geodesics are determined by the value of the angular momentum ($a$). Closer to the black hole is located the unstable circular geodesic and further the stable one.  

%%%BIBLIOGRAPHY

\end{document}